\documentclass[reprint,amsmath,amssymb,aps,prb,superscriptaddress]{revtex4-1}
\usepackage{units}
\usepackage{amsmath}
\usepackage{url}
\usepackage{comment}
\usepackage{natbib}
\usepackage{textcase}
\usepackage{amssymb}
\usepackage{graphicx}
\usepackage{bm}
\usepackage{float}
\usepackage{multirow,microtype,color,relsize,ulem}
\usepackage{braket}
\usepackage{subfigure}
\usepackage{hyperref}
\usepackage[utf8]{inputenc}
\usepackage[english]{babel}
\hypersetup{colorlinks=true,linkcolor=blue,citecolor=blue}
\hypersetup{linktocpage}

\begin{document}
	\title{Shape-preserving beam transmission through non-Hermitian disordered lattices}
	
	\author{A. F. Tzortzakakis}
	\affiliation{Physics Department, University of Crete, Heraklion, 71003, Greece}
	\author{K. G. Makris}
	\affiliation{Physics Department, University of Crete, Heraklion, 71003, Greece}
	\affiliation{Institute of Electronic Structure and Laser, FORTH, 71110 Heraklion, Crete, Greece}
	\author{S. Rotter}
	\affiliation{Institute for Theoretical Physics, Vienna University of Technology (TU Wien), A-1040 Vienna, Austria}
	\author{E. N. Economou}
	\affiliation{Physics Department, University of Crete, Heraklion, 71003, Greece}
	\affiliation{Institute of Electronic Structure and Laser, FORTH, 71110 Heraklion, Crete, Greece}

\begin{abstract}
We investigate the propagation of Gaussian beams through optical waveguide lattices characterized by correlated non-Hermitian disorder. In the framework of coupled mode theory, we demonstrate how the imaginary part of the refractive index needs to be adjusted to achieve perfect beam transmission, despite the presence of disorder. Remarkably, the effects of both diagonal and off-diagonal disorder in the waveguides and their couplings can be efficiently eliminated by our non-Hermitian design. Waveguide arrays thus provide an ideal platform for the experimental realization of non-Hermitian phenomena in the context of discrete photonics.

\end{abstract}

\date{\today}
\maketitle

\section{Introduction}
\par Wave propagation through complex disordered media is a topic of intense research interest due to its immediate physical and technological relevance. Generally speaking, the presence of disorder leads to fundamental phenomena such as multiple scattering and Anderson localization, which have been extensively studied for both quantum and classical waves \cite{loc1,loc2,loc3,loc4,loc5,loc6}. A direct manifestation of such wave scattering is the highly complex intensity pattern that is formed due to multi-path interference. With the advent of spatial light modulators and wavefront shaping techniques, interest has been growing in controlling such scattering pattern of waves propagating in complex media, for various novel applications in imaging and detection establishing the area of disordered photonics \cite{rand1,rand2,rand3,rand35}. A great challenge is to overcome the detrimental effects of multiple scattering to achieve enhanced transmission through such a complex medium of disorder. A variety of experimental methods has been recently proposed \cite{rand4,rand5,rand6,rand7,rand75,rand8,rand9}. However, most of these techniques rely on the availability of transmission resonances of the random medium and as a result require sophisticated wavefront shaping methods and adaptive imaging iteration algorithms. An alternative strategy would be to modify the scattering medium, instead of the incoming optical beam. Along these lines, one could naively expect that already the inclusion of gain inside the scattering medium will be sufficient to increase the transmission. Unfortunately, however this is not typically the case and more sophisticated methods are required to overcome and control the scattering phenomena in inhomogeneous environments.

\begin{figure}[tb]
	\centering
	\includegraphics[clip,width=1\linewidth]{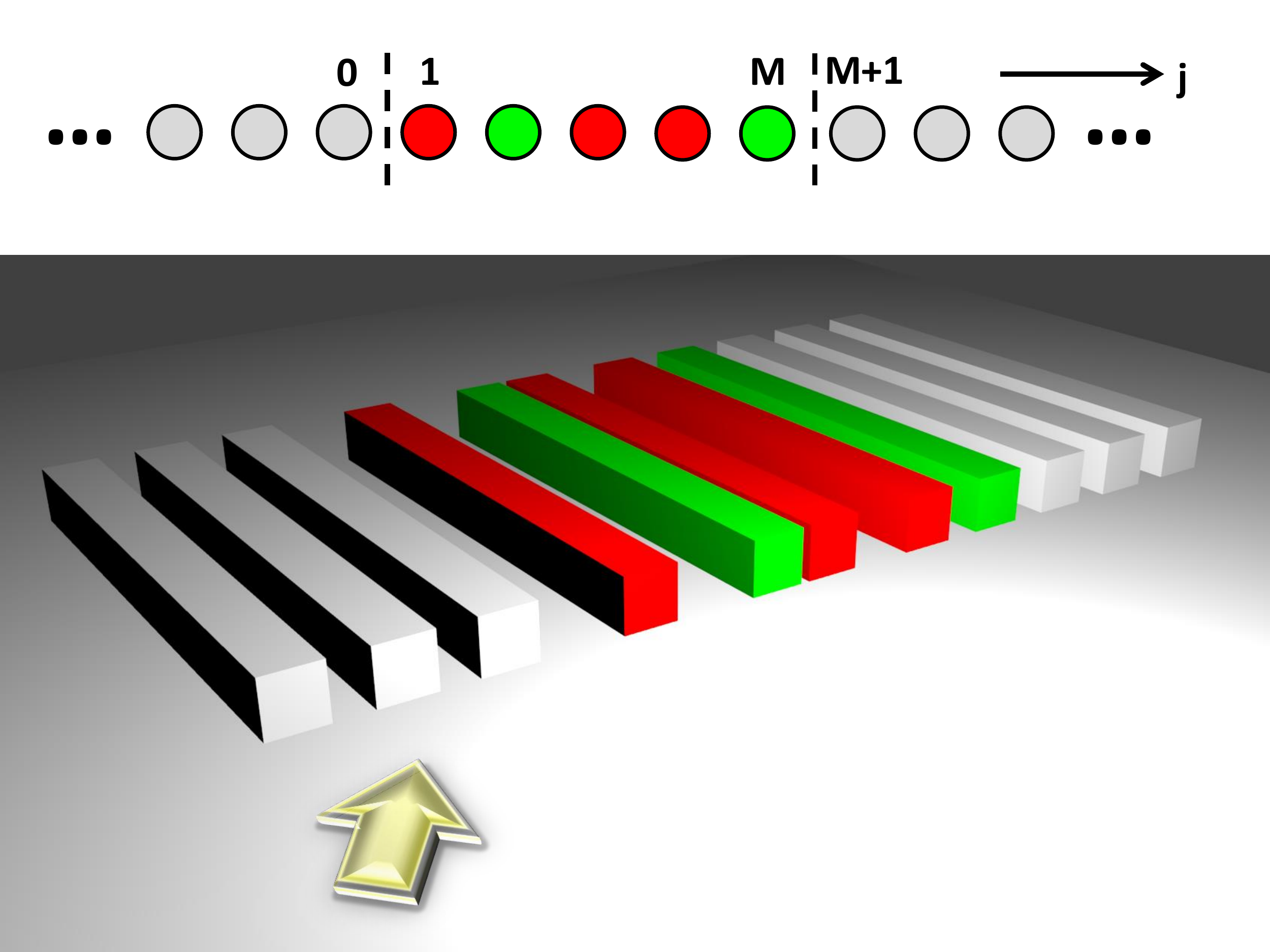}
	\caption{Schematic depiction of disordered waveguide structure. The two semi-infinite sublattices on the right and left are shown as gray channels. The disordered lattice on the middle ($1\leq j \leq M$) is illustrated by red and green colored channels for gainy and lossy waveguides, respectively. The random position of the waveguides is due to the random couplings between nearest neighbors. The arrow denotes the direction of the incident Gaussian wavepacket coupled to the left periodic array.}
	\label{sch}
\end{figure}

\par In an other direction \cite{ci1}, the study of optical structures characterized by amplification (gain) and dissipation (loss) has led to the development of a new research field, that of non-Hermitian photonics \cite{review1, review2, Pile, Gbur,review3,Longhi,review4,review5}. In particular, the introduction of the concepts of parity-time ($\mathcal{PT}$) symmetry \cite{Bender1,Bender2,Bender3} and exceptional points \cite{EP1,EP2,EP3,EP35,EP4,EP5} in optics, where gain and loss can be physically implemented \cite{PT1,PT2,PT3,PT4,PT5,PT55}, triggered a number of theoretical and experimental works, which have demonstrated the potential applications of such non-Hermitian systems. The rich behavior and novel features of these structures has led to a plethora of experimental realizations of various optical devices spanning from unidirectional invisibility to broadband wireless power transfer \cite{PT6,PT7,PT8,PT9,PT10,PT11,PT12,PT13,PT14,PT15,PT16,lin_unidirectional_2011, feng_experimental_2013, horsley_spatial_2015,konotop_families_2014,zhu_one-way_2013,antilas} and non-Hermitian Anderson localization \cite{Andreas,nhAnd,impur,nhmbl,longhi,hat}.

\par In the context of non-Hermitian photonics, it was recently demonstrated that is possible to suppress the effects of localization and thus achieve perfect transmission by considering correlated non-Hermitian disorder. In particular, one can derive a novel class of  waves that have constant intensity (CI-waves) everywhere in space, even inside the scattering area \cite{ci1,ci2}. Such waves exist in guided and scattering media with gain and loss in both one and two spatial dimensions \cite{ci3,ci4,ci45}. It was also experimentally demonstrated that CI-pressure waves are possible in the acoustical domain \cite{ci5}. The existing works focus on excitation of CI-waves by plane waves and so far there is no experimental observation in the optical domain. Therefore, in this work we will go beyond CI-waves (that have infinite extent) and will show that it is also possible to obtain reflectionless wavepackets that propagate through disordered environments in a similar fashion. More specifically, we will derive the correlated non-Hermitian disorder that is crucial for finite beams to be perfectly transmitted. The physical system that we are going to investigate is that of coupled paraxial waveguide arrays. Both disorder and gain/loss can be implemented in this type of versatile integrated platform, which is ideal for controlled optical experiments. Diagonal, as well as off-diagonal disorder, in the complex refractive index and the coupling coefficients respectively, will be systematically examined in 1+1 dimensional lattices. The robustness of such an effect and the relations of these wavepackets to CI-waves will be studied in detail.

%%%%%%%%%%%%%%%%%%%%%%%%%%%%%%%%%%%%%%%%%%%%%%%%%%%
\section{Disordered waveguide array}
\par We begin our analysis by considering optical wave
propagation in non-Hermitian disordered waveguides. This model is based on coupled mode theory and can be considered a non-Hermitian version of the Anderson tight binding model \cite{loc5}. In particular, we consider a waveguide lattice of evanescently coupled waveguides along the $x$-transverse direction. The light propagates along the $z$-longitudinal direction and is described by the following normalized paraxial coupled equations:
\begin{equation}
i\frac{\partial \psi_{j}}{\partial z}+c_{j}\psi_{j+1}+c_{j-1}\psi_{j-1}+\epsilon_{j}\psi_{j}=0
\label{par}
\end{equation}
where $\psi_{j}$,$\epsilon_{j}$ are the modal field amplitude and the propagation constant (which plays the role of the on-site energy) at the $j^{th}$ waveguide site, respectively. $c_{j}$ is the corresponding coupling coefficient between nearest neighbors (here we assume that $c_{j+1\to j}=c_{j \to j+1}\equiv c_{j}$). 
\par The total lattice consists of three different sublattices, two periodic ones in the asymptotic regions and a disordered one in the middle. More specifically, for $j<1$ and $j>M$ we assume two semi-infinite periodic sublattices, namely: 
\begin{equation}
	\epsilon_{j}=0, \quad c_{j}=1 \quad \textnormal{for} \quad j<1 \quad \textnormal{or} \quad j>M 
\end{equation}
\par In the middle region, $1\leq j \leq M$, our lattice is disordered. The geometry of the problem is graphically depicted in the schematic of Fig.~\ref{sch}. In particular, we examine two different types of disorder: (a) on-site (diagonal) disorder (paragraph V) and off-diagonal disorder on the coupling coefficients (paragraph VI). 

The main focus of our study is to examine if it is possible to suppress the transverse reflection by considering complex correlated disorder. More specifically, we are interested in understanding the effect of non-Hermiticity on the  transport of a finite wavepacket across the disordered region. When we have Hermitian disorder only ($\epsilon_{j}$ real) then most of the light is reflected in the transverse direction of the lattice and the propagation of the beam gets distorted. The question we will try to solve is whether the addition of gain and loss  in form of an imaginary part of the diagonal elements $\epsilon_j$ can remedy these detrimental effects altogether. 

%%%%c%%%%%%%%%%%%%%%%%%%%%%%%%%%%%%%%%%%%%%%%%%%%%%%
\section{Continuous limit and CI-Waves }
Before we continue to the main part of our work, it is beneficial to examine the continuous limit of our discrete problem and the connection of Eq.~(\ref{par}) to CI-waves. This investigation is going to provide us with the necessary intuition for the form of the correlated non-Hermitian disorder we have to use. For this purpose we first examine the case of diagonal disorder ($c_{j}=c=const.$), which means that disorder exists only on the waveguide channels, such that Eq.~(\ref{par}) now becomes:
\begin{equation}
i\frac{\partial \psi_{j}}{\partial z}+c(\psi_{j+1}+\psi_{j-1})+\epsilon_{j}\psi_{j}=0
\label{par1}
\end{equation}
where $\epsilon_{j}$ takes on spatially correlated random values to make the continuum limit meaningful. By applying the gauge transformation $\psi_{j}=\Psi_{j}e^{i2cz}$ and allowing $c=\frac{1}{(\Delta x)^2}$ (see \cite{ci2} for more details), the above equation in the continuum limit $(\Delta x \to0)$ can be written as:
\begin{equation}
	i\Psi_{z}+\Psi_{xx}+V(x)\Psi=0
\end{equation}
where V(x) is the disordered potential. If in the above Schr\"odinger-type equation, that describes wave propagation in the paraxial limit,  we also assume that we have a plane wave with propagation constant $k_{z}$ along the $z$-direction: $\Psi=\Phi(x) e^{ik_{z}z}$, the equation we obtain is mathematically equivalent to the 1-D Helmholtz equation.
\par It has been shown \cite{ci1} that this equation supports constant intensity solutions, if the potential satisfies the following relation:
\begin{equation}
	V(x)=[k_{x}W(x)]^2-ik_{x}W'(x) +k_{z}
	\label{cont}
\end{equation}
where $W$ is an arbitrarily chosen real, smooth function of $x$ and $W'=\frac{dW}{dx}$. In the context of integrability soliton theory these potentials naturally appear and are sometimes called Wadati potentials\cite{konotop_families_2014,wadati,horsley19,konotop_2}. In the case of Eq.~(\ref{cont}) with $x$ a continuous variable, the second order Helmholtz operator $\hat{H}$ can be factorized \cite{horsley19} as follows:
\begin{equation}
\begin{gathered}
\hat{H} \equiv -\hat{D}^2, \quad \textnormal{where} \quad \hat{D}=-i\sigma_{x}\partial_{x}+\sigma_{y}\kappa-i\sigma_{z}k_{x}W(x)
\end{gathered}
\label{hald}
\end{equation}
with $\hat{D}$ being the (first order) Dirac operator of the generalized Haldane model with imaginary mass, $\sigma_{x},\sigma_{y},\sigma_{z}$ are the usual Pauli matrices and $\kappa=\sqrt{k_{z}}$. In the above expression the Pauli matrices act on the spinor $\begin{pmatrix}
\Phi_{1}\\ \Phi_{2}
\end{pmatrix}$ , where $\Phi_{1}$ is the real and $\Phi_{2}$ the imaginary part of the total field: $\Phi\equiv \Phi_{1}+i\Phi_{2}$.
\par One can easily verify (see \cite{horsley19} for details) that $\hat{D}$ possesses a constant intensity eigenstate\cite{ci1,ci2}:
\begin{equation}
	\Psi(x,z)=\exp[ik_{z}z+ik_{x}\int_{0}^{x}W(x')dx']
	\label{cisol}
\end{equation}
 as long as the so-called degree \cite{deg} of $W$ is zero (i.e. $W$ has the same sign in $\pm \infty$).
%%%%%%%%%%%%%%%%%%%%%%%%%%%%%%%%%%%%%%%%%%%%%%%%%%%

\section{Discrete CI-Waves for Diagonal Disorder}
\par Inspired by the previous paragraph, we will now extend our study to the discrete case, by considering the realistic physical model of Fig.~\ref{sch}. Let us assume that instead of an incoming plane wave (continuous case) we have a Bloch wave that propagates from the left sublattice ($j<1$) of the form:
\begin{equation}
	\psi_{j}(z)=\exp(ik_{z}z+ik_{x} \alpha \cdot j)
\end{equation}
where $\alpha$ is the lattice constant of the two periodic sublattices, and $k_{x}$ the Bloch momentum that takes values inside the first Brillouin zone, namely $-\frac{\pi}{\alpha}\le k_{x}<\frac{\pi}{\alpha}$. The propagation constant $k_{z}$ in the two periodic sublattices, is directly related to the Bloch momentum $k_{x}$ through the dispersion relation $k_{z}=2c\cdot \textnormal{cos}(k_{x}\alpha)$, which defines the band of the lattice (in this section we assume that $c_{j}=c=1$ for simplicity). On the other hand, discretization of Eq.~(\ref{cisol}) leads us to the following ansatz for the $\psi_{j}$, which constitutes a discrete CI-wave:
\begin{equation}
	\psi_{j}(z)=\exp(ik_{z}z+ik_{x}\alpha \sum_{m=1}^{j}W_{m})
	\label{ans}
\end{equation}
Direct substitution of this ansatz into Eq.~(\ref{par}) leads us to the conclusion that we must consider a non-Hermitian potential, with a real part of the following form:
\begin{equation}
	\epsilon_{R,j}=2\cos(k_{x}\alpha)-\cos(k_{x}\alpha W_{j})-\cos(k_{x}\alpha W_{j+1})
	\label{rec}
\end{equation}
and a corresponding imaginary part:
\begin{equation}
	\epsilon_{I,j}=\sin(k_{x}\alpha W_{j})-\sin(k_{x}\alpha W_{j+1})
	\label{cipot}
\end{equation} 

From Eq.~(\ref{rec}), we can see that the potential becomes periodic if $W_{j}=1$, $\forall j$. If $W_{j}$ is random, then the potential takes on also random values around $2\cos(k_x\alpha)$. The strength of disorder can be controlled by adjusting the amplitude of $W_{j}$. 

\par Another important point is the boundary conditions on the two interfaces of the disordered region at $j=1$ and at $j=M$ (see Fig.~\ref{sch}). In order to achieve a smooth transition from one sublattice to another, the continuity of the $k_{z}$-component across the interface is essential. Thus, we need to apply the appropriate boundary conditions for the function $W$, which are the following perfect transmission boundary conditions \cite{ci3}:
\begin{equation}
		W_{1}=W_{M}=1
	\label{bc}
\end{equation}
We also need to emphasize that the above boundary conditions ensure both that the degree of $W$ is zero \cite{horsley19,deg} and that the average of gain and loss is zero, $\sum_{j=1}^{M}\epsilon_{I,j}=0$ (mean reality condition\cite{ci3}). 

%%%%%%%%%%%%%%%%%%%%%%%%%%%%%%%%%%%%%%%%%%%%%%%

\section{Wadati Wavepackets for Diagonal Disorder}

\par In this paragraph we are going to investigate the main question of our work, which is how to achieve perfect and shape-preserving transmission of an incoming beam through a discrete disordered medium. Our strategy is based on the concept of discrete CI-waves that was described in the previous paragraph. However, in the present work and for the sake of being realistic we are employing  a Gaussian beam in space (or, equivalently a Gaussian wavepacket in time) instead of a pure Bloch wave. This beam/wavepacket has a central wavenumber corresponding to the discrete CI-wave and propagates from the left to the right, starting from the left periodic sublattice. The width of such a beam is denoted with $\sigma$, and its center is located around some waveguide with index $j_{0}<1$ and has a specific group velocity. As $\sigma$ tends to infinity the pure Bloch wave is recaptured. In other words our initial beam can be expressed as:
\begin{equation}
\begin{gathered}
	\psi_{j}(z=0)=\exp[ik_{x}\alpha(j-j_{0})-(\frac{j-j_{0}}{\sigma})^2] 
\end{gathered}
\label{beam}
\end{equation}
Inside the disordered region, we seek finite, constant-width propagating wavepackets of the (approximate) form:
\begin{equation}
\begin{gathered}
|\psi_{j}(z)|=\exp[-(\frac{j-j_{0}(z)}{\sigma})^2],\\
\quad \textnormal{with} \quad 	j_{0}(z)=j_{0}+2\sin(k_{x}\alpha)z 
\end{gathered}
	\label{ci}
\end{equation}
Since these type of beams exist only for the discrete version of non-Hermitian Wadati potentials, we call these solutions ``Wadati wavepackets".

\par The price we pay for considering a Gaussian beam as our initial condition is an extra limitation. In particular, the whole analysis of the constant intensity waves (Eqs.~\ref{hald}) is based on an incident plane wave (or Bloch wave in our case)  which corresponds to $\sigma$ equal to infinity instead of the finite $\sigma$ of our Gaussian beam. In other words, the potential we have introduced is designed for a single wavenumber $k_{x}$ while our beam is composed of a large number of different wavenumbers. The components of the beam which correspond to $k'_{x}\ne k_{x}$ will then be scattered due to the randomness of the potential and distort the pattern of the wavepacket. This effect will be sharpened if we increase the amount of the potential's randomness. As a result, in order for this distortion to be weak enough for our solution to be of the expected form, $W$ cannot be a totally random function, but a ``slowly'' varying one. In other words, the jumps from one site to another should not be arbitrarily large, but rather satisfy: $\Delta W_{j}=|W_{j+1}-W_{j}| \sim \alpha << \sigma$, $\forall j $. Of course, in the limit of large $\sigma$, this limitation ceases, as our incident beam now becomes a Bloch wave.
\begin{figure}[tb]
	\centering
	\subfigure{\includegraphics[clip,width=0.495\linewidth]{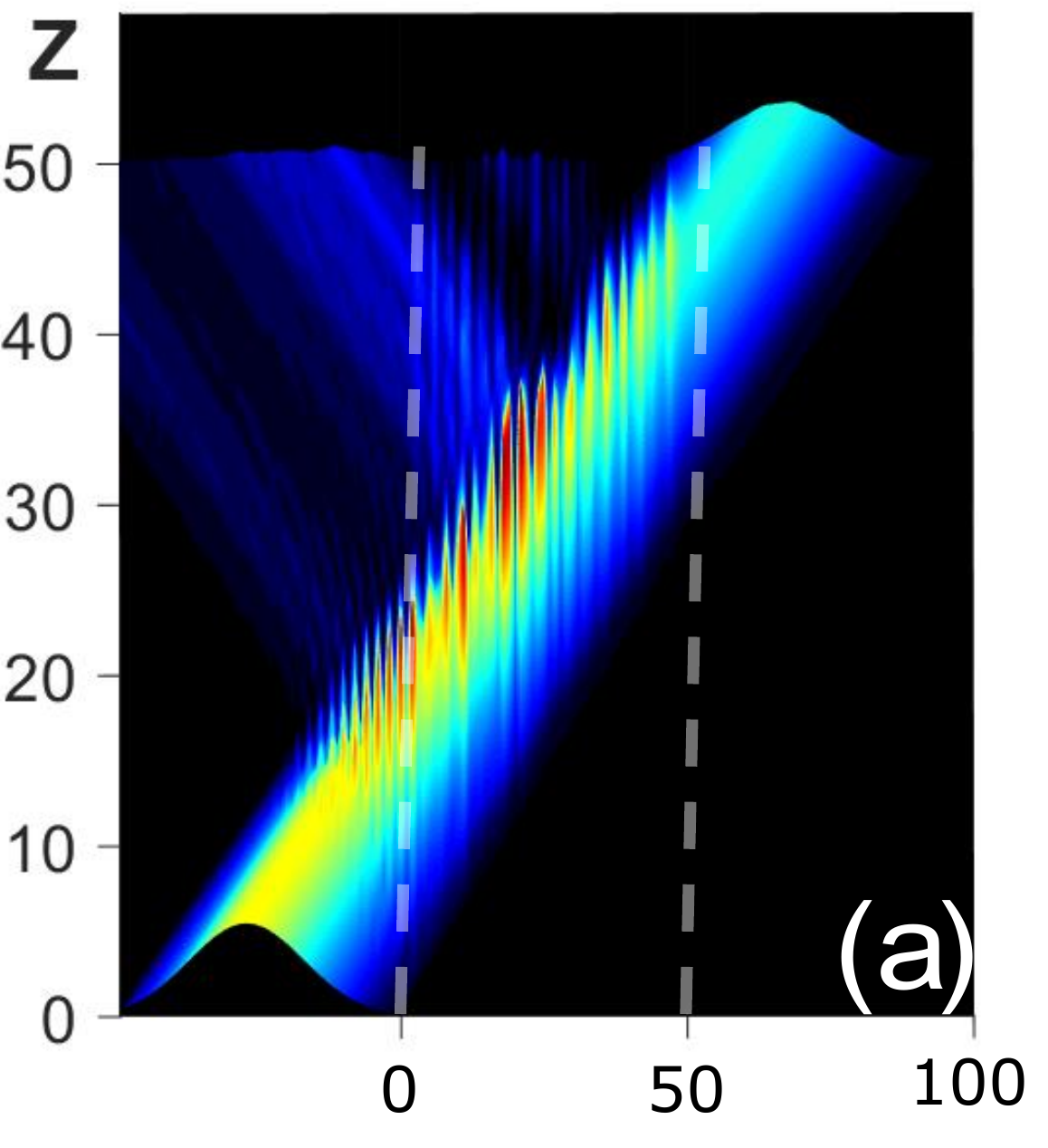}}\subfigure{\includegraphics[clip,width=0.5\linewidth]{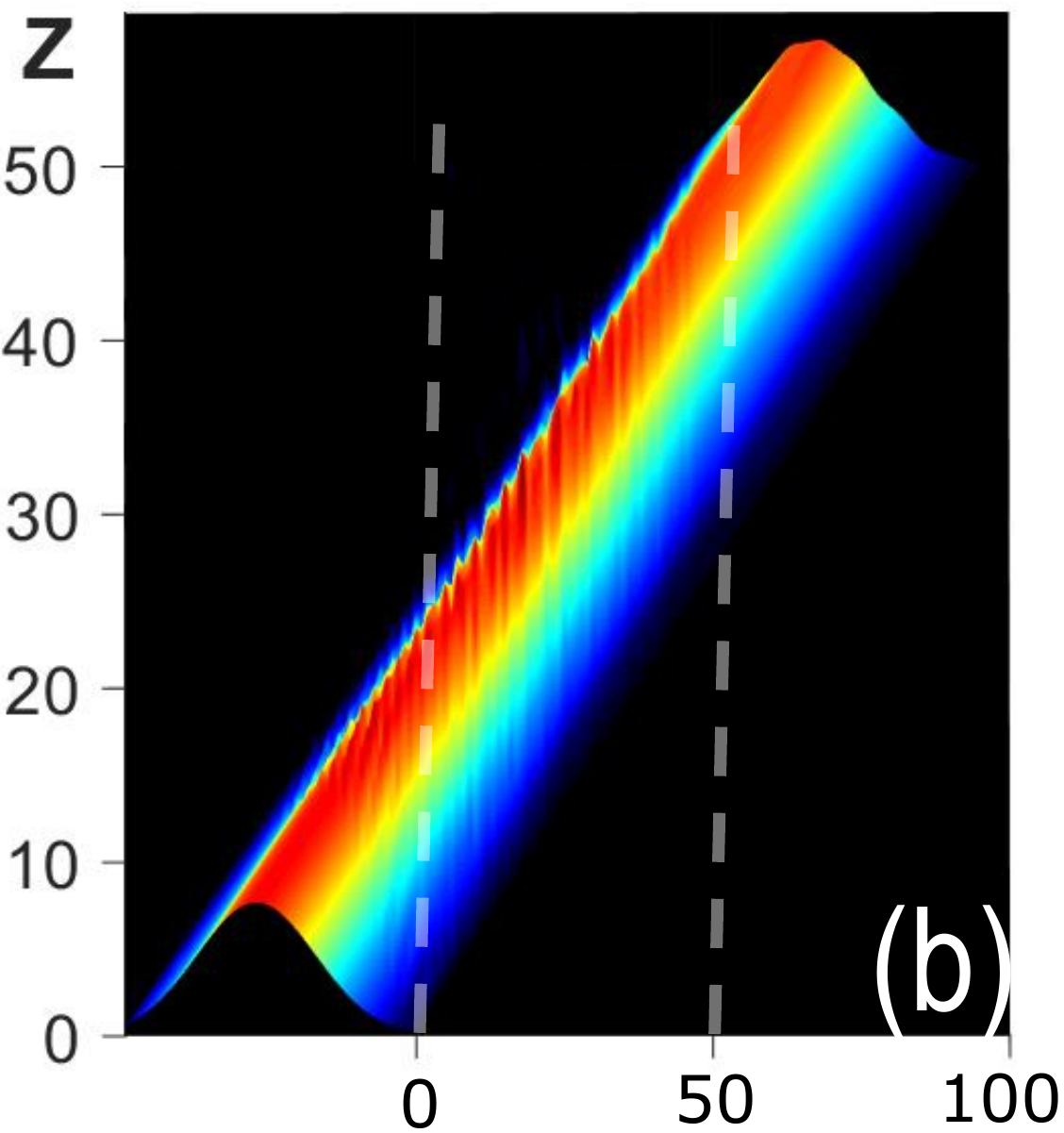}}
	\subfigure{\includegraphics[clip,width=0.5\linewidth]{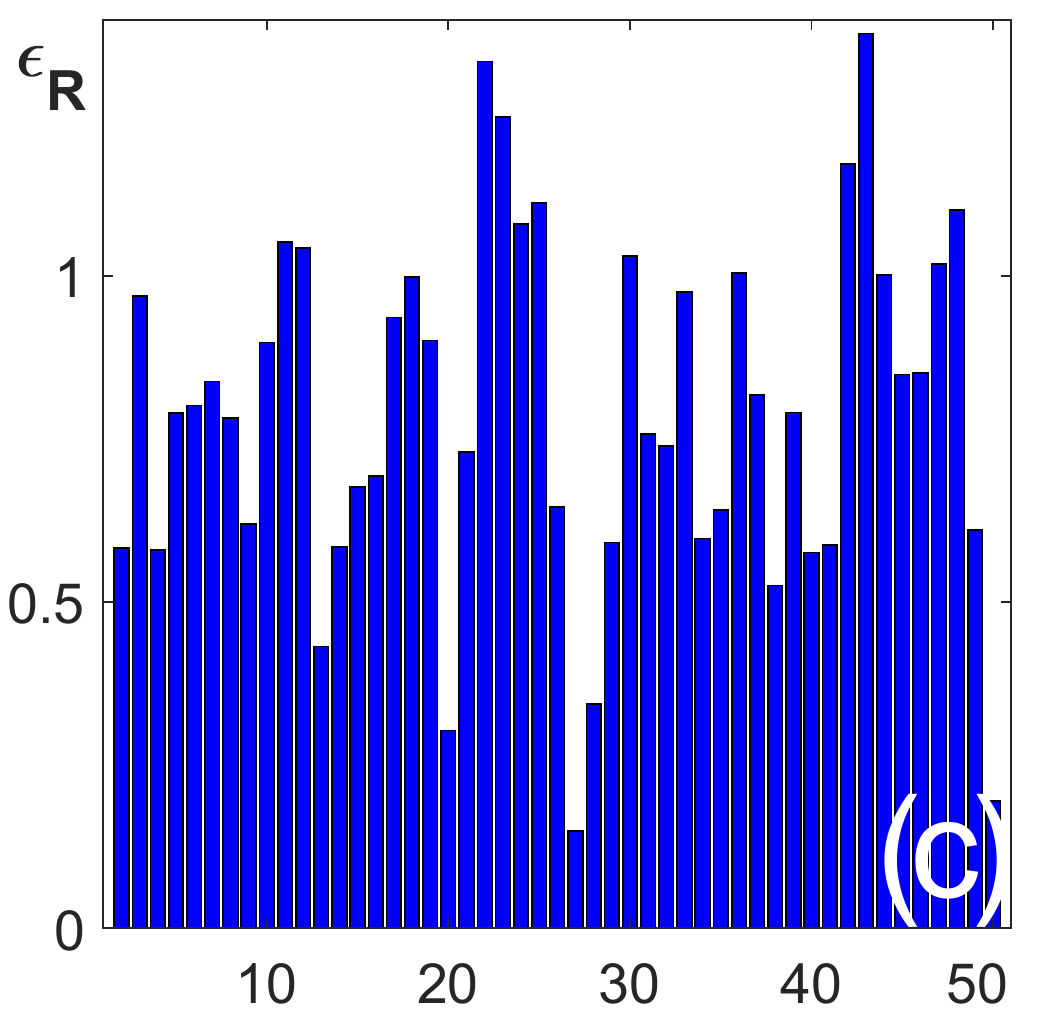}}\subfigure{\includegraphics[clip,width=0.503\linewidth]{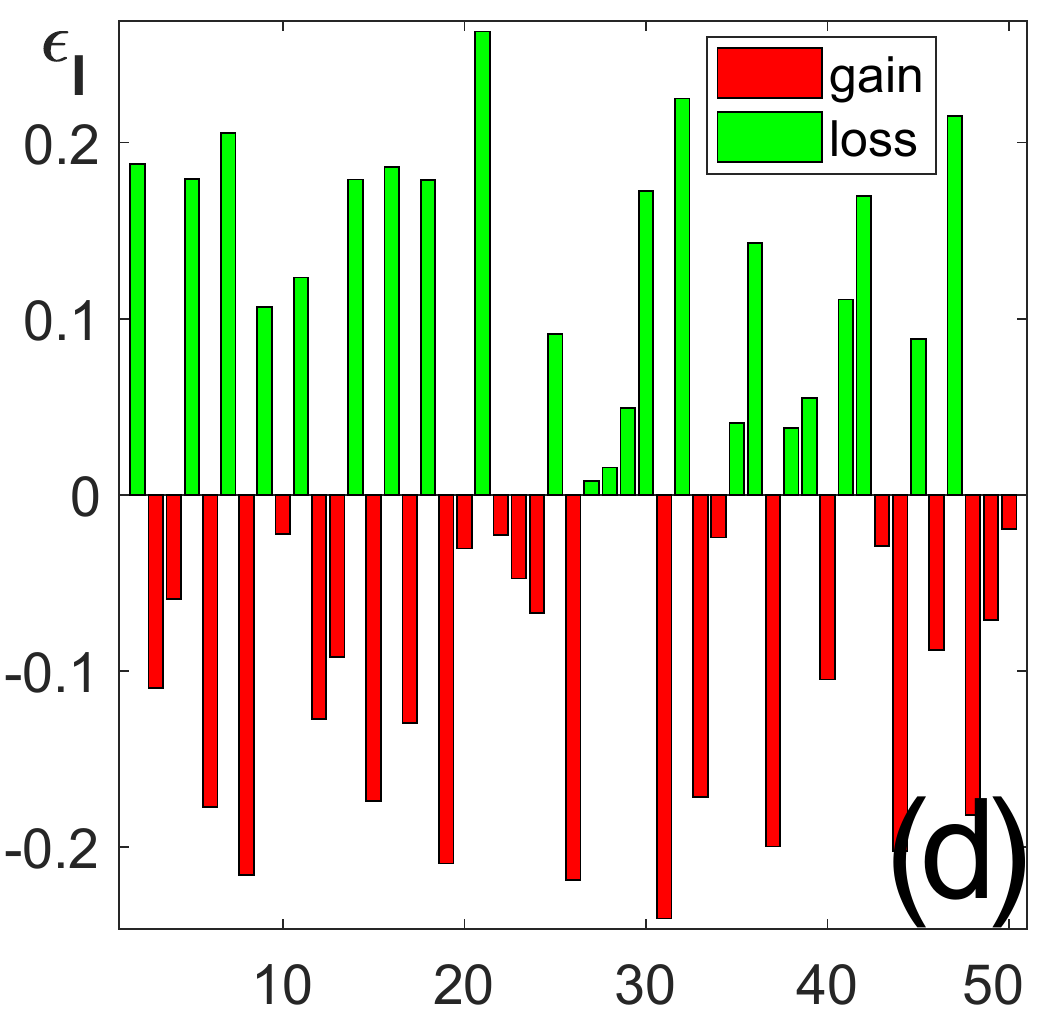}}
	\caption{(a) Intensity of a beam propagating in a potential with real disorder (reflection occurs). (b) Same as in (a) but with imaginary part based on relation (\ref{cipot}) (reflection is minimized). Dashed lines denote the interfaces among the sublattices. (c) Real part of the potential. (d) The corresponding imaginary part. In all the figures above, the $x$-axis represents the waveguide number $j$. Here we have set: $M=50$, $j_{0}=-37.5$ and $\sigma=30$.}
	\label{diag}
\end{figure}
\par Let us now examine propagation of beams through Wadati potentials (Fig.~\ref{diag}). We initially consider the Gaussian beam of Eq.~(\ref{beam}) impinging on a random Hermitian potential $\epsilon_{R}$ of Eq.~(\ref{rec}) (Fig.~\ref{diag}(a)) and then include the appropriate imaginary part based on Eq.~(\ref{cipot}) (Fig.~\ref{diag}(b)). In Fig.~\ref{diag}(c),(d) the corresponding real and imaginary parts of the potential are depicted, satisfying the above smoothness condition
\par As one can see, in the Hermitian case the reflection due to disorder is very strong leading to very low transmission. On the contrary, for the non-Hermitian case the transmission is almost perfect and the Wadati beam/wavepacket maintains its transverse form for every value of the propagation distance. Thus by adding the appropriate imaginary part to the real (random) potential, the beam penetrates the disordered region and propagates with (almost) constant peak amplitude. As expected for a finite Gaussian wavepacket, it also spreads in its width during propagation.
\par We accentuate here that this shape-preserving perfect transmission of the Wadati wavepacket is unidirectional. This means that an incident beam from the right sublattice (with $k'_{x}=\pi +k_{x}$) does not lead us the same results, since the time-reversal symmetry of the lattice is broken (due to non-Hermiticity). The transmittance from the right incidence is again one, as our system is reciprocal, but we also get strong reflection. In order to get the same shape-preserving transmission from the right, we would have  conjugate potential $\epsilon_{j} \to \epsilon_{j}^{*}$ when injecting from the opposite side.

\begin{figure}[tb]
	\centering
	\includegraphics[clip,width=0.7\linewidth]{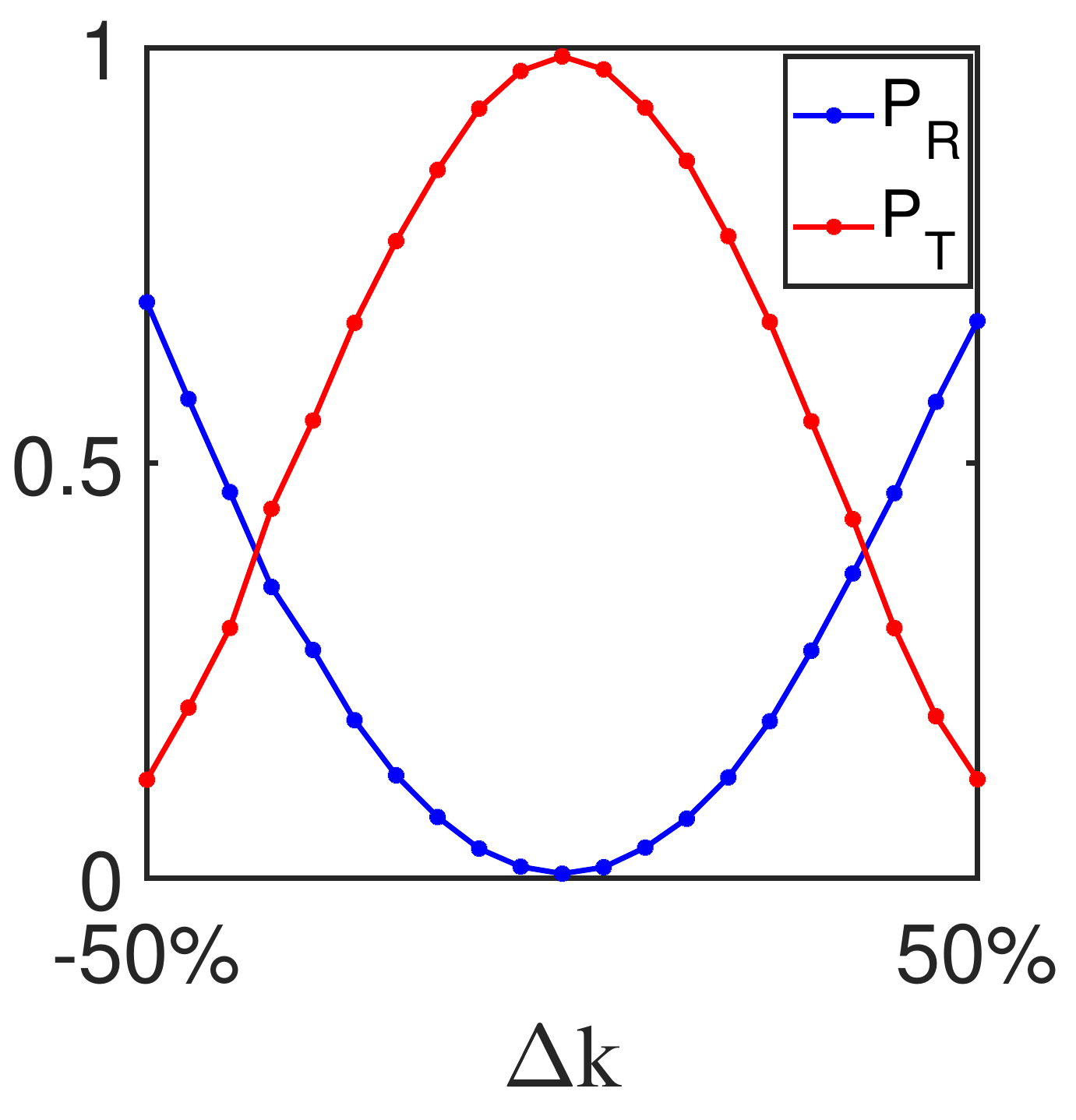}
	\caption{Power of the transmitted ($\textnormal{P}_{T}$, red line) and reflected ($\textnormal{P}_{R}$, blue line) wave, divided by the power of the initial beam, (Eq.~(\ref{int})), as a function of the percentage deviation (detuning) from the required wavelength value $k_{x}$: $\Delta k=k'_{x}-k_{x}$. In this case we have set $k_{x}=\frac{\pi}{2}$, for the case of diagonal disorder. An averaging over 500 realizations of disorder has been performed for these results.}
	\label{sens}
\end{figure}

\par As we have mentioned before, the non-Hermitian potential is by default designed to support a discrete CI-wave at a single transverse wavenumber $k_{x}$. Therefore an important question is how sensitive is the transmission of the corresponding Wadati wavepacket to changes of its central wavenumber. For this reason, we calculate the power $(\textnormal{P}=\sum_{j}|\psi_{j}|^2)$ transmitted to the right sublattice $\textnormal{P}_{T}$, as well as the corresponding reflected power $\textnormal{P}_{R}$, over the power of the input beam, after the passing of the beam through the disordered region:
\begin{equation}
	\textnormal{P}_{T}=\dfrac{\textnormal{P}_{Transmitted}}{\textnormal{P}_{Incident}}=\dfrac{\sum_{j=M+1}^{\infty}|\psi_{j}|^2}{\textnormal{P}_{Incident}}  \label{intt}
\end{equation}
\begin{equation}
	\textnormal{P}_{R}=\dfrac{\textnormal{P}_{Reflected}}{\textnormal{P}_{Incident}}=\dfrac{\sum_{j=-\infty}^{M}|\psi_{j}|^2}{\textnormal{P}_{Incident}}
	\label{int}
\end{equation} 
as a function of the percentage deviation $\Delta k \%$ between the required wavenumber $k_{x}$ and the beam's wavenumber $k'_{x}$: $\Delta k=k'_{x}-k_{x}$. The results are shown in Fig.~\ref{sens} and are averaged over 500 realizations of disorder. Both $\textnormal{P}_{T}$ and $\textnormal{P}_{R}$ exhibit a parabolic behavior, with the peak (dip) located at the expected value of $\Delta k=0$. In addition, we have to note that we get $\textnormal{P}_{T}\approx 1$ for a wavenumber variation $|\Delta k|\leq 10\%$; a result very close to the corresponding one from the continuous case \cite{ci3}.
\par In order to have a better physical perspective of our problem, we provide here some indicative order of magnitude estimation of the actual scales for a possible experiment. In particular, the wavelength of the beam is $\lambda_{0}\approx1 \mu m$, the distance between neighboring waveguides is $D\approx10 \mu m$, while the propagation distance $z$ is normalized over $2k_{0}n_{0}D^2$, with $n_{0}\approx3.5$ being the background refractive index of the waveguides and $k_{0}=\frac{2\pi}{\lambda_{0}}$. Finally, the potential is $\epsilon=2k^2_{0}n_{0}D^2(\Delta n_{R}+i\Delta n_{I})$, where $\Delta n$ represents the variation of the waveguide's refractive index with respect to the background value of $n_{0}$. Under these conditions, the maximum variation in the real part of the index of refraction (Fig.~\ref{diag}(c)) is approximately $\Delta n_{R}^{max}\approx 10^{-3}$ and the maximum gain (loss) used (Fig.~\ref{diag}(d)) is $g_{max}\approx3 cm^{-1}$.

%%%%%%%%%%%%%%%%%%%%%%%%%%%%%%%%%%%%%%%%%%%%%%%%%%%
%%%%%%%%%%%%%%%%%%%%%%%%%%%%%%%%%%%%%%%%%%%%%%%%%%%

\section{OFF-DIAGONAL DISORDER}
\par In this section we will examine whether it is possible to obtain a perfectly transmitting wavepacket for the case of random real couplings $c_{j}$, as encountered when the distance between neighboring waveguides is not the same. Since for this problem the discrete Wadati potential of the previous paragraph does not provide a straightforward solution, a new approach is required. 
\par Substituting the ansatz of Eq.~(\ref{ans}) in Eq.~(\ref{par}) once again, with the coupling coefficients $c_{j}$ being random this time, the obtained potential reads as follows:
\begin{equation}
\begin{gathered}
	\epsilon_{R,j}=2\cos(k_{x}\alpha)-\cos(k_{x}\alpha)(c_{j}+c_{j-1})\\
	\epsilon_{I,j}=\sin(k_{x}\alpha)(c_{j}-c_{j-1})
\end{gathered}
	\label{ciboth}
\end{equation}

\par Our result implies that, in order to cope with the randomness in the coupling matrix elements $c_{j}$, we also need to introduce complex randomness in the potential. The potential given by the relations (\ref{rec}) and (\ref{cipot}) is then modified as follows: while we need to set $W_{j}=1$, a factor involving the $c_{j}$ must be incorporated in the $\cos(k_{x}\alpha)$ and $\sin(k_{x}\alpha)$ terms of $\epsilon_{R}$ and $\epsilon_{I}$ respectively.
\begin{figure}[tb]
	\centering
	\subfigure{\includegraphics[clip,width=0.5\linewidth]{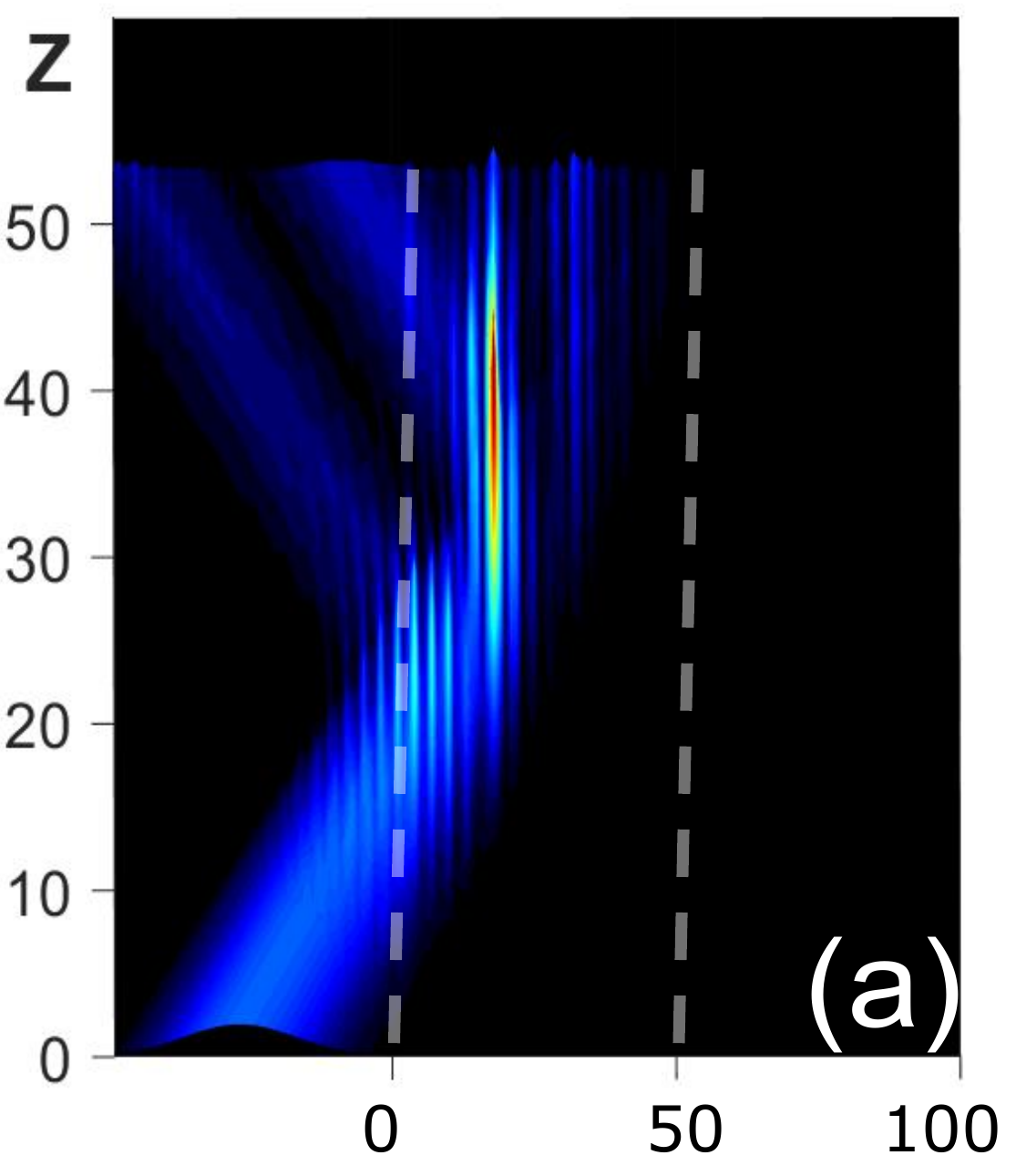}}\subfigure{\includegraphics[clip,width=0.5\linewidth]{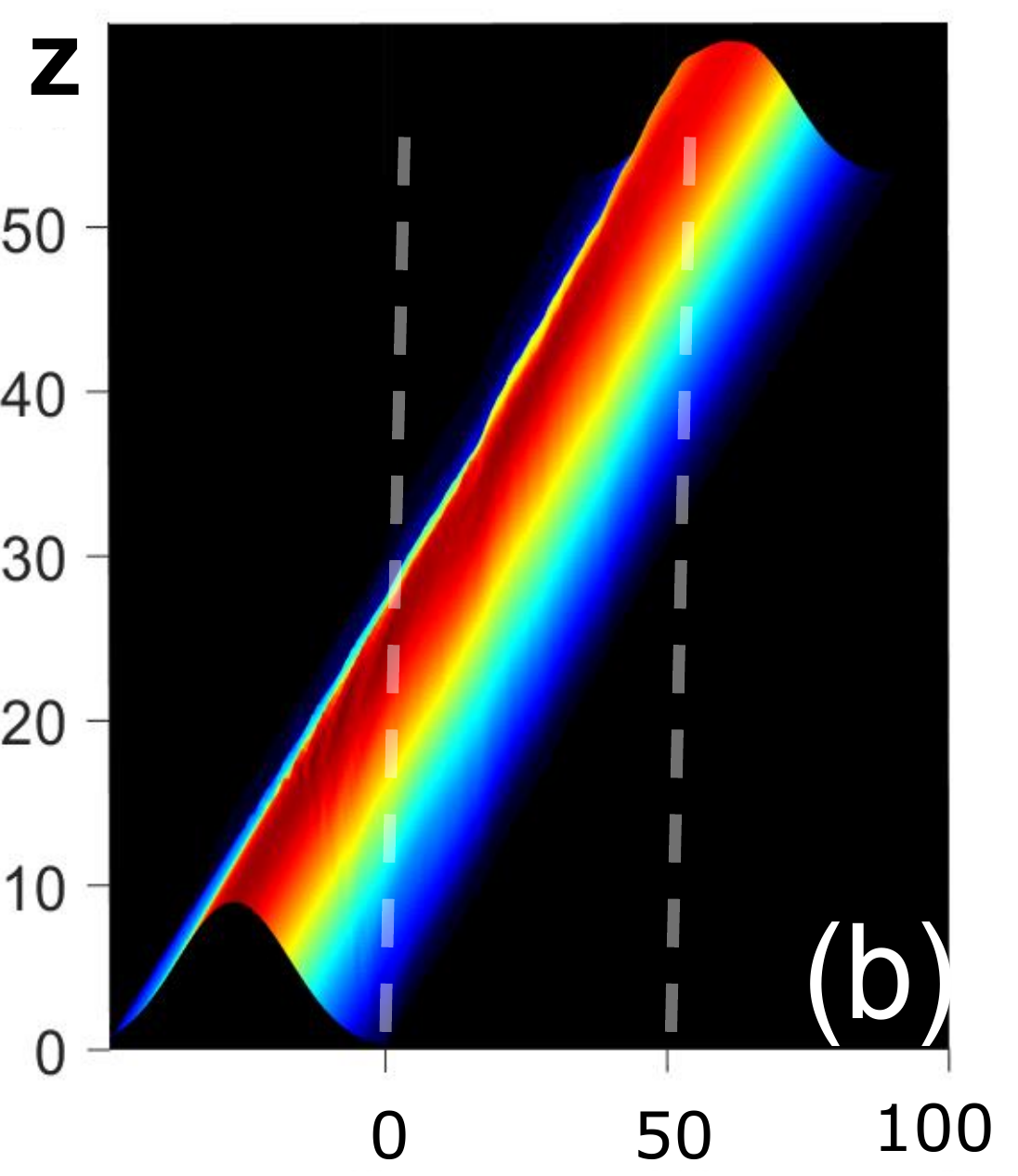}}
	\subfigure{\includegraphics[clip,width=0.33\linewidth]{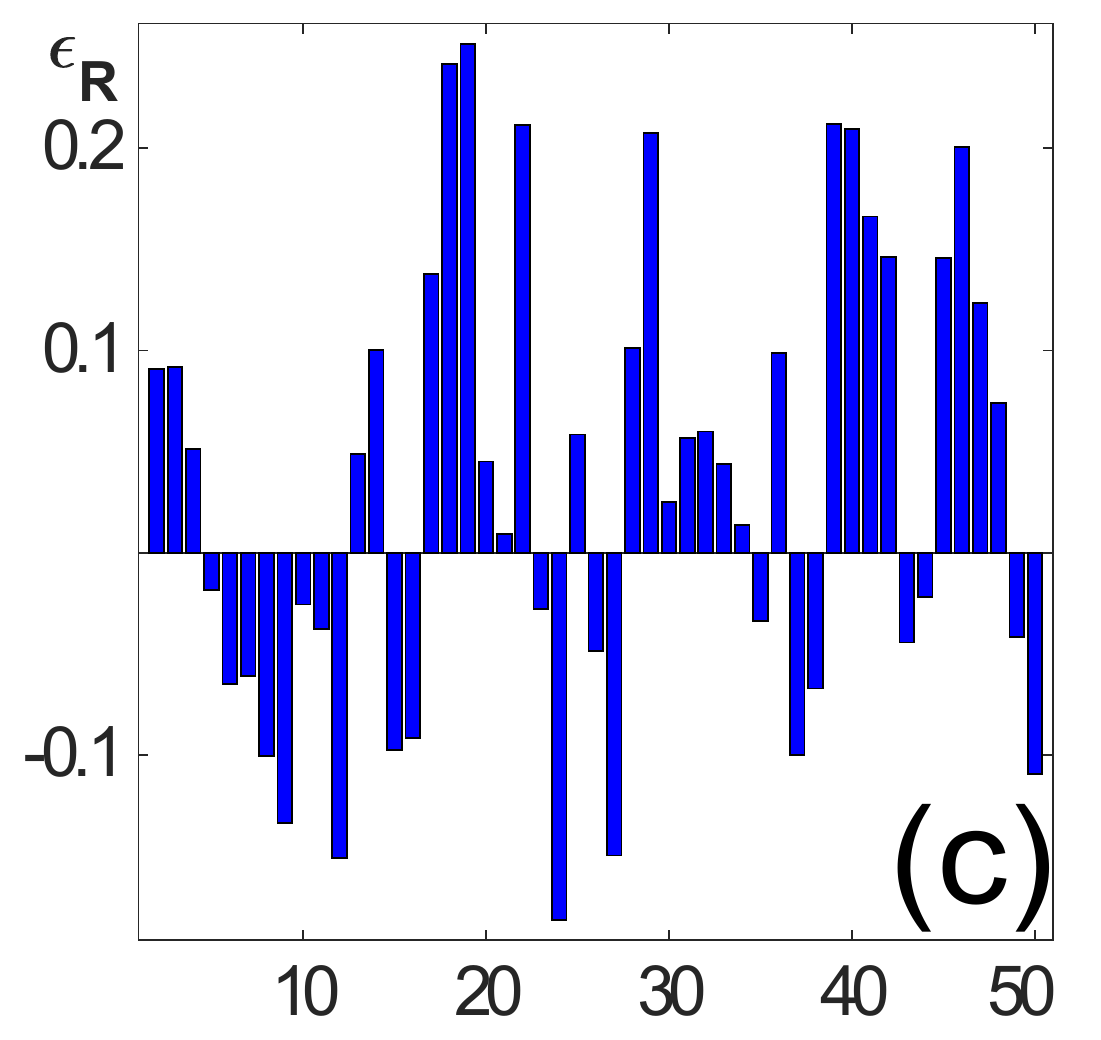}}\subfigure{\includegraphics[clip,width=0.33\linewidth]{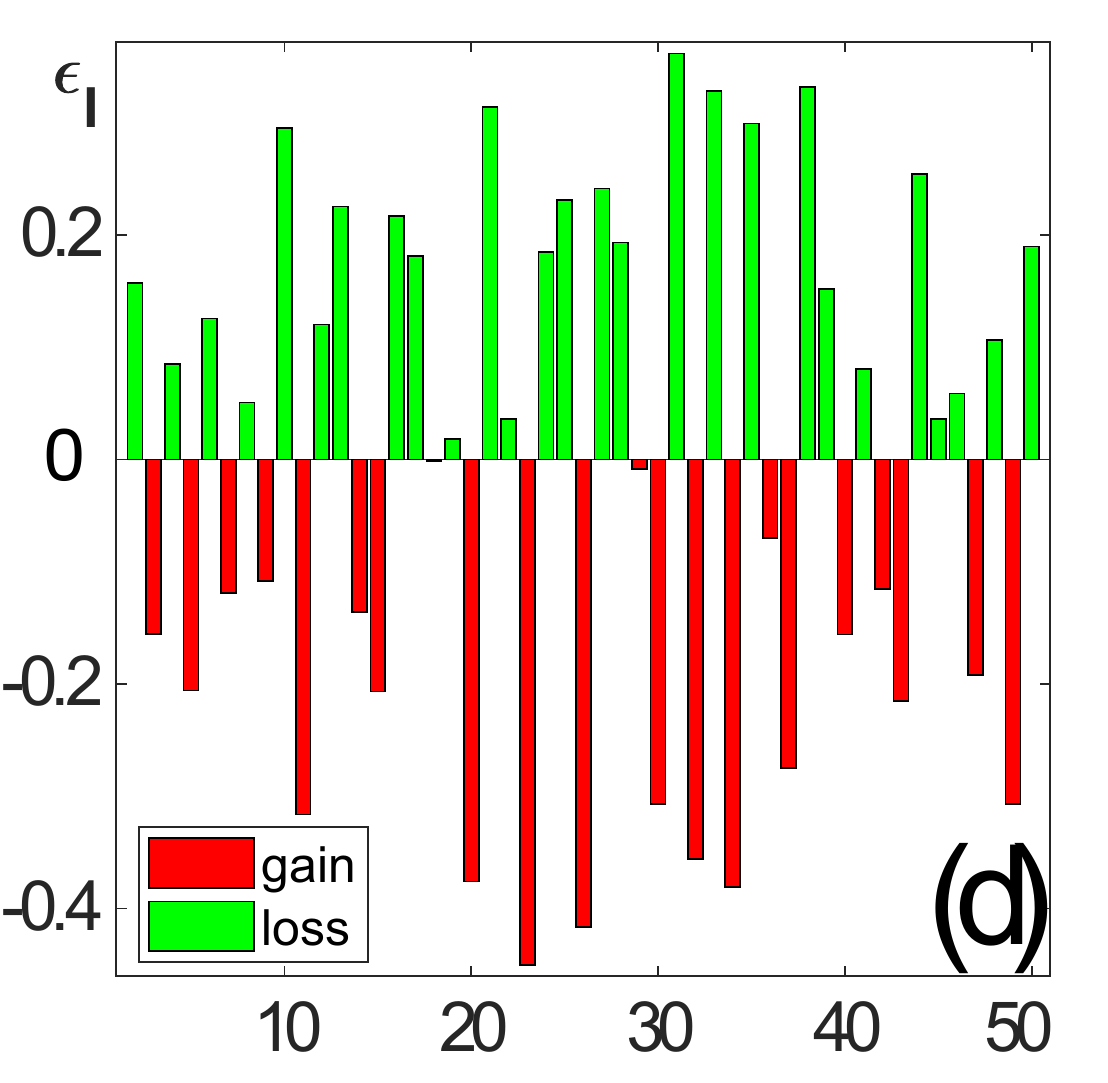}}\subfigure{\includegraphics[clip,width=0.33\linewidth]{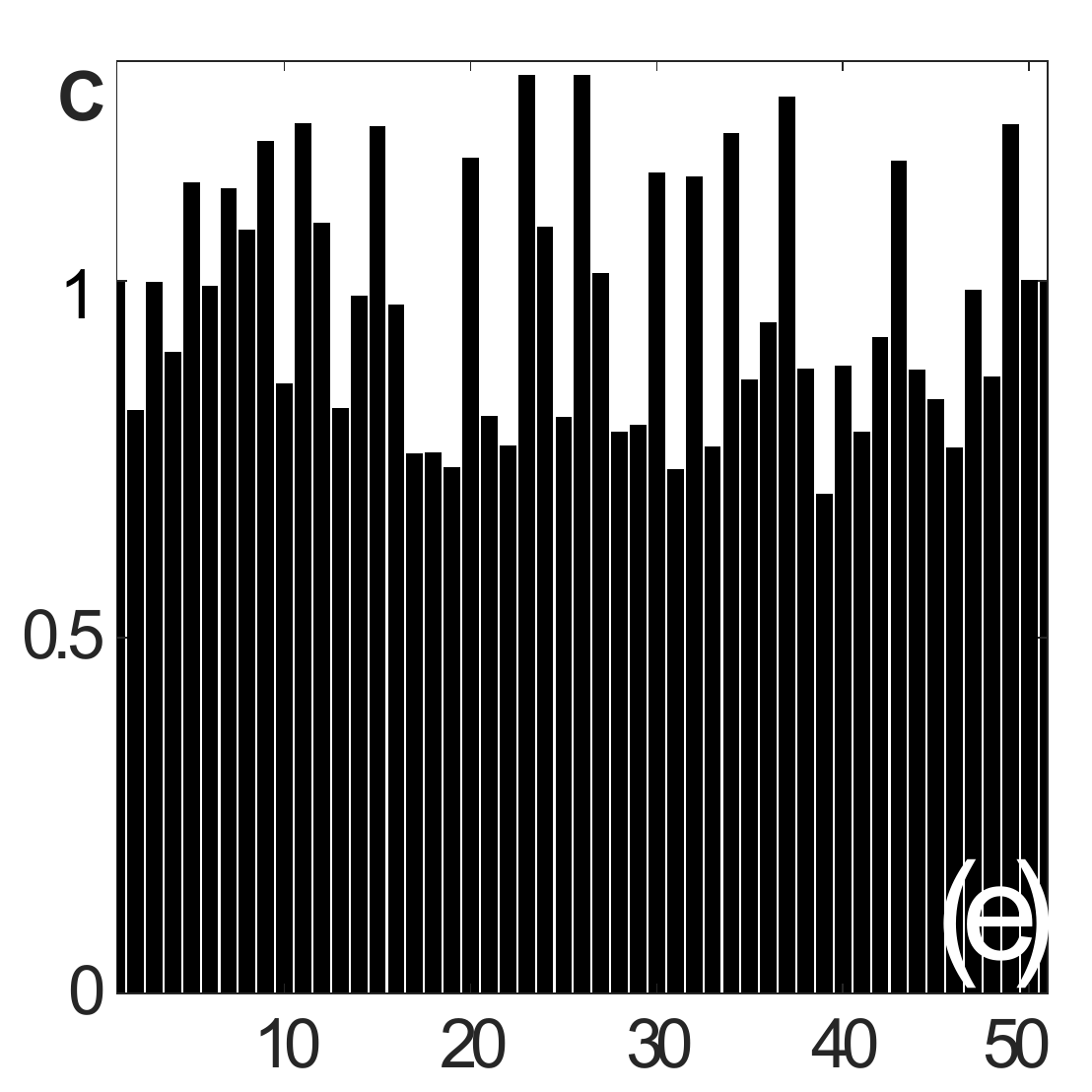}}
	\caption{(a) Intensity of a beam propagating in a lattice with random real couplings. (b) Same as in (a) but with the potential of the form of Eq.~(\ref{ciboth}) which supports constant intensity solutions. Dashed lines denote the interfaces among the sublattices. Real (c) and imaginary (d) part of the potential and coupling distribution (e) for the results shown above. In all the figures above the $x$-axis represents the waveguide number $j$. For these graphs we have set: $M=50$, $j_{0}=-37.5$ and $\sigma=30$.}
	\label{bpm3}
\end{figure}

\par The propagation of a Gaussian beam across this discrete non-Hermitian potential landscape is depicted in Fig.~\ref{bpm3}. In particular, we see in  Fig.~\ref{bpm3}(a) that the strong reflection due to disorder leads to almost zero transmission. By considering the appropriate (complex) refractive index modulation the transmission becomes perfect and shape-preserving, with almost zero reflection, as is demonstrated in Fig.~\ref{bpm3}(b) for one particular realization of disorder.
\par The coupling coefficient distribution, as well as the corresponding complex potential, are also shown in Fig.~\ref{bpm3}. We point out here, that this case of off-diagonal disorder seems to be more robust than the case of diagonal disorder, meaning that the reflection is even more insignificant than in the results shown in Fig.~\ref{diag}.
\begin{figure}[tb]
	\centering
	\includegraphics[clip,width=0.7\linewidth]{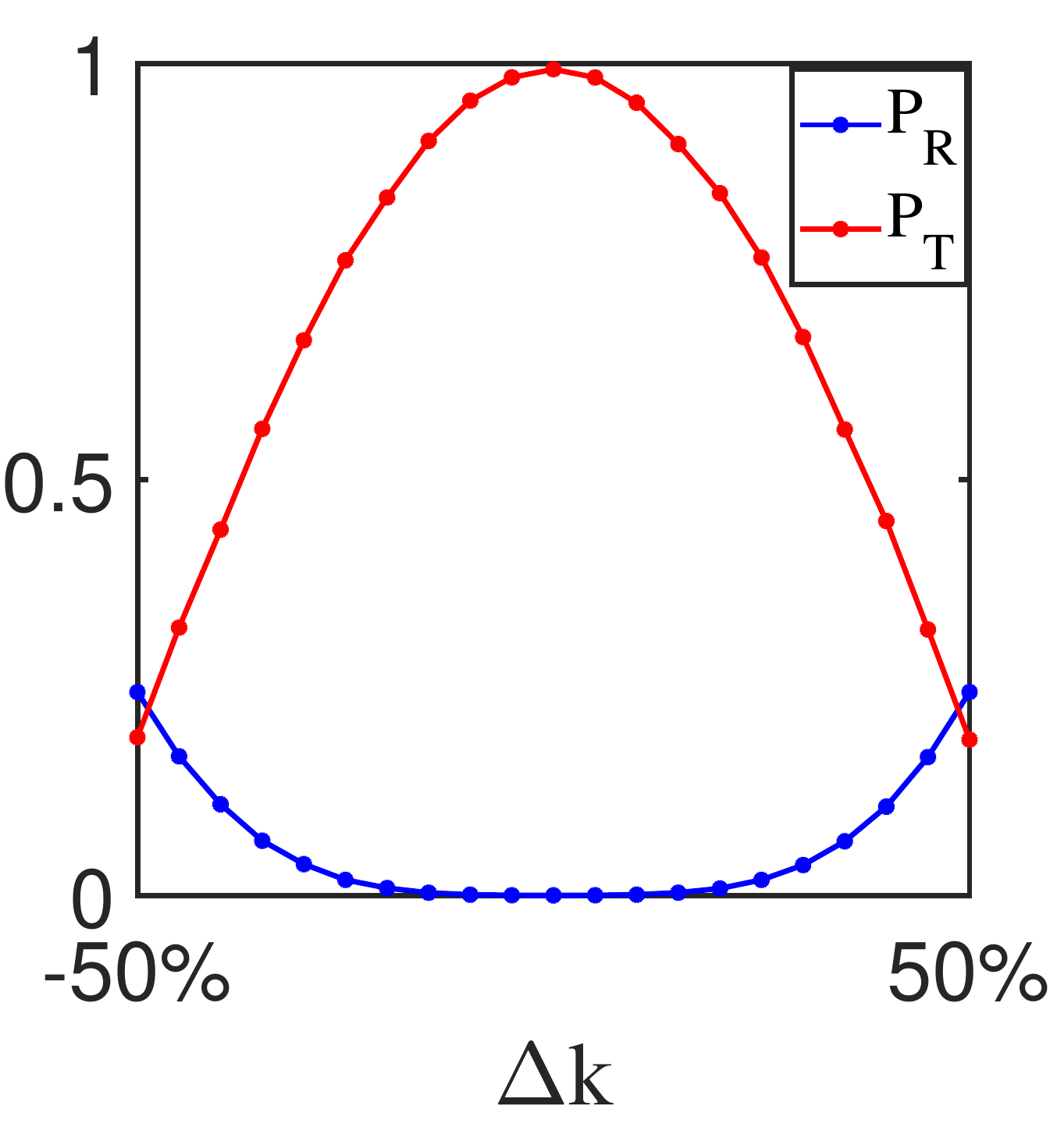}
	\caption{Power of the transmitted ($\textnormal{P}_{T}$, red line) and reflected ($\textnormal{P}_{R}$, blue line) wave, divided by the power of the initial beam (Eq.~(\ref{int})), as a function of the percentage deviation (detuning) from the required wavelength value $k_{x}$: $\Delta k=k'_{x}-k_{x}$. In this case we have set $k_{x}=\frac{\pi}{2}$, for the case of off-diagonal disorder. An averaging over 500 realizations of disorder has been performed for these results.}
	\label{sensoffd}
\end{figure}
\par Finally, in Fig.~\ref{sensoffd} we plot the transmitted and reflected power, defined by Eqs.~(\ref{intt},\ref{int}), as a function of the wavelength detuning: $\Delta k=k'_{x}-k_{x}$. As in Fig.~\ref{sens}, $\textnormal{P}_{R}$ and $\textnormal{P}_{T}$ exhibit again a parabolic behavior. However, here the perfect transmission peak is broadened: $\textnormal{P}_{T}\simeq1$ and $\textnormal{P}_{R}\simeq0$ for $|\Delta k| \leq 20\%$. In addition, the reflected power, contrary to the diagonal disorder case, reaches values up to $0.3$. We attribute this behavior to the trapping of the beam in lossy regions of the lattice, leading to a rapid decay in the beam's intensity.

\section{Discussion and Conclusions}
In this paper we have proposed a systematic methodology to eliminate reflection due to disorder in realistic discrete systems consisting of coupled waveguides. Our strategy is based on an extension of the recently introduced concept of CI-waves to realistic discrete systems. In particular, we have studied the perfect transmission of Gaussian wavepackets through random optical lattices in 1+1 dimensions, which are non-Hermitian, due to the complex index of refraction. In the Hermitian limit, or even when the lattice has only loss or only gain elements, the transmission is low and the field is strongly distorted. However, for non-Hermitian disorder, where the real and the imaginary parts are correlated in the way we describe, almost perfect transmission is achieved. Such an enhanced transmission, despite the strong transverse reflection due to Anderson localization, is based on the extension of CI-waves in the discrete domain. Two different cases of on-diagonal (Wadati wavepackets) and off-diagonal disorder have been thoroughly examined and for both cases a near-perfect and shape-preserving transmission of an incoming Gaussian wavepacket is observed. We believe that this systematic study will pave the way for the direct experimental realization of CI-waves to integrated photonic waveguide structures. Also extensions of this concept to lasers and coherent perfect absorbers in disordered waveguide lattices should be within immediate reach.  

\section{Acknowledgements}
We acknowledge support from the European Commission under project NHQWAVE (Grant No. MSCA-RISE 691209) and  under European Commission project Visorsurf  (grant agreement 736876). S.R. is supported by the Austrian Science Fund (FWF) through project P32300 (WAVELAND).

\bibliographystyle{longbibliography}

\end{document}